\documentclass[
aps,%
12pt,%
final,%
notitlepage,%
oneside,%
onecolumn,%
nobibnotes,%
nofootinbib,%
superscriptaddress,%
noshowpacs,%
centertags]%
{revtex4}

\usepackage[utf8]{inputenc}
\usepackage[T1]{fontenc}
\usepackage[dvips]{graphicx}
\usepackage{amsmath,amssymb}
\usepackage{mathptmx}
\usepackage{booktabs}
\usepackage{bm}
\usepackage[dvips]{color}
\usepackage{float}
\usepackage[dvipdfmx]{hyperref}
\usepackage{lipsum}

\usepackage[labelfont=bf,labelsep=period,justification=raggedright,singlelinecheck=false]{caption}

\begin{document}
\renewcommand{\thefootnote}{\fnsymbol{footnote}}

\title{Monopole and Instanton Effects on Connected and Disconnected Correlations for Scalar Density}
  
\author{{Masayasu Hasegawa$^{\left.1\right), *}$} \\ $^{\left.1\right)}$Bogoliubov Laboratory of Theoretical Physics, Joint Institute for Nuclear Research, Dubna, Russia \\ (Received March 2, 2023; revised March 2, 2023; accepted March 2, 2023)}

\footnote[0]{\noindent$^{*}$Electronic address: hasegawa@theor.jinr.ru}

\begin{abstract}
  {\bf Abstract--}This study investigates the effects on the connected and disconnected correlations for the scalar density that are induced by created monopoles and instantons in the QCD vacuum. To reveal the effects, we add a monopole and anti-monopole pair in the gauge field configurations in \textit{SU}(3) by applying the monopole creation operator to the vacuum. We vary the magnetic charges of the monopole and anti-monopole and increase the number of monopoles and anti-monopoles in the configurations. The Dirac operator of overlap fermions preserves the exact chiral symmetry in lattice gauge theory and exact zero-modes exist in its spectrum. The eigenvalues and eigenvectors of the overlap Dirac operator have been calculated using these configurations, and the numbers of instantons and anti-instantons which are created by these additional monopoles and anti-monopoles have been estimated from the numbers of topological charges in our previous studies. In this study, we demonstrate the preliminary results that instantons and monopoles influence the masses that are evaluated from the connected and disconnected correlation functions for the scalar density using low-lying eigenvalues and eigenvectors of the overlap Dirac operator.
\end{abstract}

\maketitle

\section{\textnormal{Introduction}}

Monopole condensation causes color confinement, as explained by 't Hooft and Mandelstam~\cite{tHooft2,Mandelstam1} and instantons induce chiral symmetry breaking, as demonstrated by the instanton vacuum~\cite{Dyakonov6} and instanton liquid model~\cite{Shuryak2}. Color confinement and chiral symmetry breaking are closely tied to one another through monopoles and instantons in the QCD vacuum. However, it is difficult to reveal the quantitative relations and effects of the instantons and monopoles on observables by perturbative calculations because of the strong interaction in the low-energy region of the QCD. Therefore, we perform simulations of lattice gauge theory and investigate the effects of monopoles and instantons on hadron spectroscopy.

In our research, we apply the monopole creation operator to the vacuum and add the monopoles and anti-monopoles to the gauge field configurations of the quenched approximation in \textit{SU}(3)~\cite{Bonati2}. We vary the magnetic charges of the monopole creation operator to increase the number of monopoles and anti-monopoles.

We then calculate the eigenvalues and eigenvectors of the overlap Dirac operator that preserves the exact chiral symmetry in lattice gauge theory~\cite{Ginsparg1, Neuberger1, Neuberger2, Lusher1, Chandrasekharan1}, using the normal configurations and the configurations to which the monopoles and anti-monopoles are added. The overlap fermions have exact zero modes in their spectra; therefore, we estimate the number of instantons and anti-instantons that are created by the additional monopoles and anti-monopoles from the number of topological charges. Finally, we investigate the impacts induced by monopole and instanton creations on hadron spectroscopy.

We have found that the added monopoles and anti-monopoles create the instantons and anti-instantons and quantitatively evaluated how many monopoles and anti-monopoles create the instantons and anti-instantons~\cite{DiGH3}.

At finite temperatures, we quantitatively demonstrated that the length of the long monopole loops lengthens and the transition temperature of color deconfinement rises by increasing the values of the magnetic charges of the additional monopoles and anti-monopoles~\cite{Hasegawa4}.

We have found that the effects that are caused by the instantons and anti-instantons created by the additional monopoles and anti-monopoles on observables~\cite{Hasegawa2} are as follows: the value of the renormalized chiral condensate in $\overline{\text{MS}}$-scheme at 2 GeV which is an order parameter of the chiral symmetry breaking and is defined as negative values decreases. The renormalized light quark masses in $\overline{\text{MS}}$-scheme at 2 GeV become heavy. The decay constants and masses of pion and kaon become heavy. The catalytic effect in which the lifetime of the charged pion becomes shorter than the experimental result. We have demonstrated that the finite lattice volume and the discretization do not affect these outcomes~\cite{Hasegawa4}.

Furthermore, we have estimated the instanton effects on the eta-prime meson mass~\cite{Hasegawa3} from two different estimation methods as follows: (i) calculations of the connected and disconnected correlation functions for the pseudoscalar density~\cite{DeGrand3}. (ii) Estimations using the Witten--Veneziano mass formula and the numerical results of the topological susceptibility and pion decay constant~\cite{Giusti8}. We have quantitatively demonstrated that the eta-prime meson mass becomes heavy with increasing the number density of the instantons and anti-instantons.

In this report, we calculate the connected and disconnected correlation functions for the scalar density using the low-lying eigenvalues and eigenvectors of the overlap Dirac operator and estimate the instanton and monopole effects on the scalar meson, especially, considering the contributions of the disconnected correlation.

The contents of this report are as follows: in Section~\ref{sec:2}, we first briefly explain simulation parameters. We then calculate the connected and disconnected correlation functions for the scalar density using the eigenvalues and eigenvectors of the overlap Dirac operator. Finally, we estimate the monopole and instanton effects on these correlation functions and their masses. In Section~\ref{sec:3}, we provide the summary and conclusion.

This report is a contribution to the proceedings of ``the 6th International Conference on Particle Physics and Astrophysics,'' from the 29th of November 2022 to the 2nd of December 2022 at the Hotel Intourist Kolomenskoe in Moscow, Russia. This report demonstrates the preliminary results.

\section{\textnormal{Correlation functions for scalar density}}\label{sec:2}

In this section, we calculate the disconnected and connected correlation functions for scalar density, evaluate their masses, and investigate the impacts of the monopole and instanton creations on the masses.

\subsection{\textnormal{\textit{Simulation Parameters}}}

The simulation parameters are the same as our previous studies~\cite{DiGH3,Hasegawa2,Hasegawa4}. The details of the additional monopoles and anti-monopoles in the gauge field configurations and the computations of the overlap Dirac operator are explained in reference~\cite{Hasegawa2}. Here, we briefly review them.

The lattice volume is $V$ = $18^{3}\times32$ ($V_{s}\times T$). The value of a parameter $\beta$ for the lattice spacing $a$ is $\beta$ = 6.0522 and the lattice spacing is $a$ = 8.527$\times10^{-2}$ fm which is estimated from the analytical functions in~\cite{Necco1}.
\begin{table*}[htbp]
  \begin{center}
    \begin{small}
    \caption{The outcomes of $N_{I}$, $\rho_{I}^{\frac{1}{2}}$, and $\rho_{I}^{\frac{1}{4}}$ and the number of configurations $N_{\text{conf}}$ in~\cite{Hasegawa3}}\label{tb:1}
    \begin{tabular}{|c|c|c|c|c|}\hline
      $m_{c}$ & $N_{I}$ & $\rho_{I}^{\frac{1}{2}}$, GeV$^{2}$ & $\rho_{I}^{\frac{1}{4}}$, MeV & $N_{\text{conf}}$ \\ \hline
      Normal conf &  9.7(5)  & 3.85(9)$\times10^{-2}$  &  1.96(2)$\times10^{2}$ & 800 \\ \hline  
                2 &  13.6(7) & 4.57(12)$\times10^{-2}$ &  2.14(3)$\times10^{2}$ & 810 \\ \hline  
                4 &  15.7(8) & 4.92(12)$\times10^{-2}$ &  2.22(3)$\times10^{2}$ & 868 \\ \hline  
                6 &  17.7(9) & 5.22(13)$\times10^{-2}$ &  2.28(3)$\times10^{2}$ & 870 \\ \hline      
    \end{tabular}
    \end{small}
  \end{center}
\end{table*}

We add a monopole and anti-monopole pair to the configurations in \textit{SU}(3) by applying the monopole creation operator~\cite{Bonati2} to the vacuum. In this study, the magnetic charges $m_{c}$ of the monopoles vary from 2, 4, to 6 and the magnetic charges $-m_{c}$ of the anti-monopoles vary from -2, -4, to -6. The total magnetic charges of the additional monopoles and anti-monopoles that are added to the configurations are set to zero. We increase the number of monopoles and anti-monopoles in the configurations by varying the magnetic charges of the additional monopoles and anti-monopoles. Hereafter, the magnetic charge $m_{c}$ indicates that both magnetic charges are added. We prepare the normal configuration and the configurations to which the monopoles and anti-monopoles are added by varying their magnetic charges.

First, we calculate the eigenvalues and eigenvectors of the Dirac operator of the overlap fermions~\cite{Ginsparg1, Neuberger1, Neuberger2, Lusher1, Chandrasekharan1} which preserves the exact chiral symmetry in the continuum limit using these configurations. We estimate the number of instantons and anti-instantons $N_{I}$ and their number density $\rho_{I} = \frac{N_{I}}{V}$ from the topological charges $Q$ as follows: $N_{I} = \langle Q^{2} \rangle$. The topological charge $Q$ is defined from the difference between the number of zeromodes of the plus chirality $n_{+}$ and the number of zeromodes of the minus chirality $n_{-}$ as follows: $Q = n_{+} - n_{-}$.

The numerical results of $N_{I}$, $\rho_{I}^{\frac{1}{2}}$, and $\rho_{I}^{\frac{1}{4}}$ and the number of configurations $N_{\text{conf}}$ have been provided in~\cite{Hasegawa2}; however, we list the outcomes in Table~\ref{tb:1} for convenience.

\subsection{\textnormal{\textit{Connected and Disconnected Correlation Functions}}}\label{sec:2_1}

In this study, the quark propagator is defined using the eigenvalues and eigenvectors of the overlap Dirac operator as follows:
\begin{equation}
  S(\textbf{y}, y^{0}; \textbf{x}, x^{0}) \equiv \sum_{i}\frac{\psi_{i}(\textbf{x}, x^{0}) \psi_{i}^{\dagger}(\textbf{y}, y^{0})}{\lambda_{i}^{\text{mass}}},
\end{equation}
$\psi$ is the eigenvector of the massless Dirac operator. The eigenvalue $\lambda_{\text{mass}}$ is the eigenvalue of the Dirac operator with a mass term $m_{q}$ and it is defined using the eigenvalues of the massless Dirac operator $\lambda$ as follows:
\begin{equation}
\lambda_{\text{mass}} = \left(1-\frac{am_{q}}{2\rho} \right)\lambda_{i} + m_{q}\label{eq:mass_d},
\end{equation}
$\rho$ is a parameter that is set to $\rho$ = 1.4. The mass term $m_{q}$ is an input parameter. In this study, we set it to $am_{q}$ = 1.729$\times10^{-2}$ in the lattice unit. This value is approximately 40 MeV in the physical unit.

The quark bilinear operator for the scalar density is defined as follows:
\begin{equation}  
 \mathcal{O}_{S} = \bar{\psi}_{1}\left( 1 - \frac{a}{2\rho}{\mathcal{D}} \right) \psi_{2}, \ \ \mathcal{O}_{S}^{c} = \bar{\psi}_{2}\left( 1 - \frac{a}{2\rho}\mathcal{D} \right) \psi_{1}.
\end{equation}
We calculate the following connected correlation function $C_{SS}$ for the scalar density:
\begin{equation}
  C_{SS}(\Delta t) = \frac{a^{3}}{V} \sum_{\textbf{x}_{1}}\sum_{\textbf{x}_{2}, \ t}\left\langle \mathcal{O}_{S}^{c}(\textbf{x}_{2}, t) \mathcal{O}_{S}(\textbf{x}_{1}, t + \Delta t)\right\rangle\label{eq:corre_conne}.
\end{equation}
Similarly, the disconnected correlation function $C_{\text{dis-}SS}$ for the scalar density is calculated as follows:
\begin{equation}
  C_{\text{dis-}SS}(\Delta t) = \frac{a^{3}}{V} \sum_{t}\left\langle \sum_{\textbf{x}_{2}}\mathcal{O}_{S}^{c}(\textbf{x}_{2}, t) \sum_{\textbf{x}_{1}}\mathcal{O}_{S}(\textbf{x}_{1}, t + \Delta t)\right\rangle\label{eq:corre_disconne}.
\end{equation}
In our study, to reduce the statistical errors, we calculate the correlation functions from all $\textbf{x}_{1}$ to all $\textbf{x}_{2}$ and take a summation regarding the temporal direction $t/a$.
\begin{figure*}[htbp]
  \begin{center}
    \includegraphics[width=165mm]{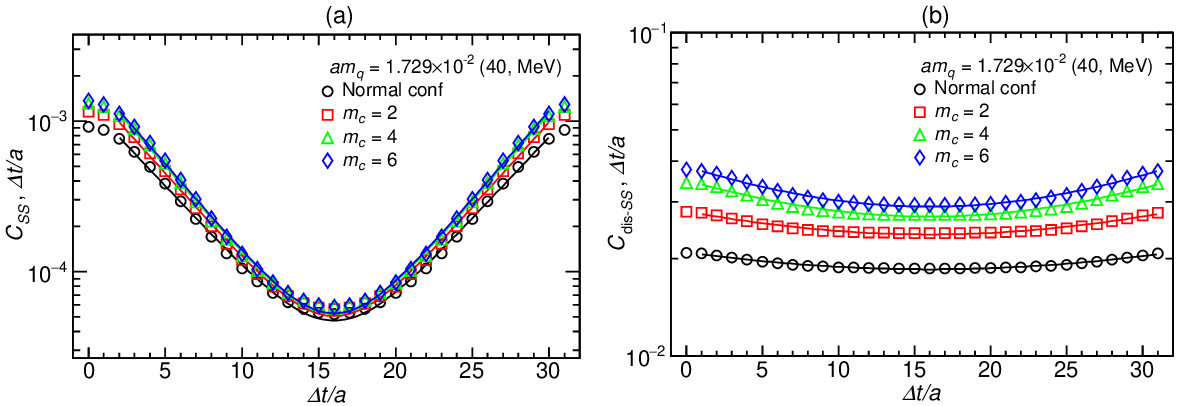}
  \end{center}
  \setlength\abovecaptionskip{-1pt}
  \caption{(a) the numerical results of the connected correlation $C_{SS}$. (b) the numerical results of the disconnected correlation $C_{\text{dis-}SS}$. The colored lines indicate the fitting results.}\label{fig:1}
\end{figure*}
\begin{table*}[htbp]
  \begin{center}
    \begin{small}
         \caption{The fitting results of $a^{4}Z$, $am$, and $\chi^{2}/\text{dof}$ and estimated masses $m$ of the connected correlation $C_{SS}$ and disconnected correlation $C_{\text{dis-}SS}$}\label{tb:2}
    \begin{tabular}{|c|c|c|c|c|c|c|}\hline
      Corre. & $m_{c}$& $a^{4}Z$ & $am$ & $m$, MeV & $FR$, $\Delta t/a$ & $\chi^{2}/\text{dof}$ \\ \hline
      $C_{SS}$ & Normal conf & 6.46(15)$\times10^{-4}$ & 0.250(2) & 5.78(5)$\times10^{2}$  & 2--30 & 20/27 \\ \cline{2-7}
                        & 2 & 8.7(2)$\times10^{-4}$ & 0.264(2) &  6.05(5)$\times10^{2}$ & 2--30 & 23/27 \\ \cline{2-7}
                        & 4 & 1.04(2)$\times10^{-3}$ & 0.269(2) &  6.22(5)$\times10^{2}$ & 2--30 & 37/27 \\ \cline{2-7}
                        & 6 & 1.09(2)$\times10^{-3}$ & 0.271(2) &  6.27(5)$\times10^{2}$ & 2--30 & 28/27 \\ \hline
    $C_{\text{dis-}SS}$ & Normal conf &  9.6(6)$\times10^{-4}$ &  3.13(13)$\times10^{-2}$     &  73(3)    & 1--31 & 3/29\\ \cline{2-7}
                                  & 2 &  1.58(8)$\times10^{-3}$ &    3.68(13)$\times10^{-2}$ &    85(3)  & 1--31 & 6/29 \\ \cline{2-7}
                                  & 4 &  2.67(10)$\times10^{-3}$ &    4.68(11)$\times10^{-2}$ &    1.08(2)$\times10^{2}$ & 1--31 & 12/29 \\ \cline{2-7}
                                  & 6 &  3.18(11)$\times10^{-3}$ &    4.96(10)$\times10^{-2}$&     1.15(2)$\times10^{2}$ & 1--31 & 8/29 \\ \hline
    \end{tabular}
    \end{small}
  \end{center}
\end{table*}

We fit the following curve $F$ with two parameters $m$ and $a^{4}Z$ to the numerical result of the correlation functions $C_{SS}$ and $C_{\text{dis-}SS}$ as shown in Fig.~\ref{fig:1}:
\begin{equation}
  F(t) = \frac{a^{4}Z}{am}\exp\left(-\frac{T m}{2}\right)\cosh\left\{m\left(\frac{T}{2} - t\right)\right\}\label{eq:fitc}.
\end{equation}
The fitting results $m$ and $a^{4}Z$ and numerical results of the masses $m$ that are estimated from the connected and disconnected correlations for the scalar density are listed in Table~\ref{tb:2}.

\begin{figure*}[htbp]
  \begin{center}
    \includegraphics[width=165mm]{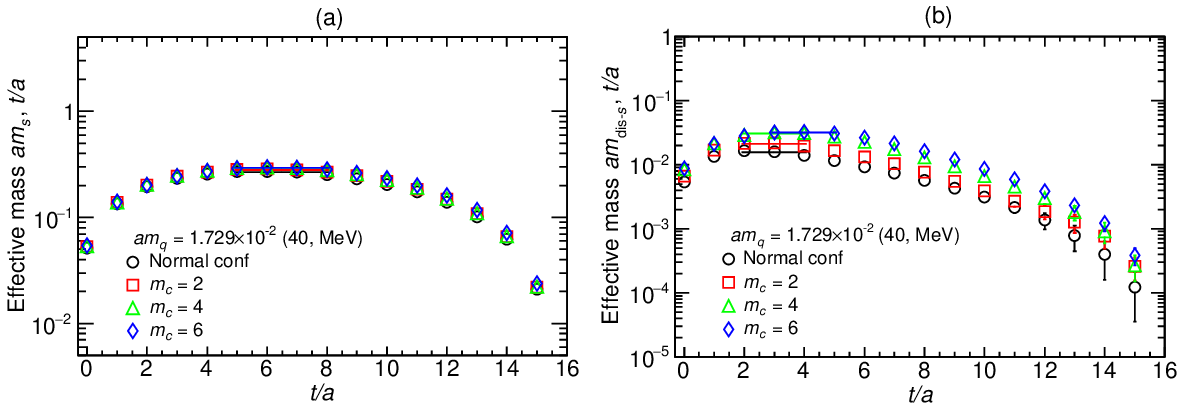}
  \end{center}
  \setlength\abovecaptionskip{-1pt}
  \caption{(a) the effective mass $am_{s}$ calculated from the connected correlations. (b) the effective mass $am_{\text{dis-}s}$ calculated from the disconnected correlations. The colored lines indicate the fitting results.}\label{fig:2}
\end{figure*}

Next, we calculate the effective mass $am_{\text{eff.}}$ using the numerical results of the correlation functions~(\ref{eq:corre_conne}) and~(\ref{eq:corre_disconne}), which is defined as follows:
\begin{equation}
  am_{\text{eff.}}(t) = -\log \frac{C(t + 1)}{C(t)}\label{eq:effemass}.
\end{equation}
We then fit a constant function $y = A_{\text{eff.}}$ to the numerical results of the effective masses $am_{\text{eff.}}$ as shown in Fig.~\ref{fig:2}. The fitting results are listed in Table~\ref{tb:3}. We set the fitting range to be narrow because the values of $\chi^{2}/\text{dof}$ become large when the fitting ranges become larger.

However, the right panel of Fig.~\ref{fig:2} shows that the numerical results of the effective mass of the disconnected correlation gradually decrease and do not achieve the plateau even if the temporal direction of the lattice is sufficiently large; therefore, we suppose that the mass estimated from the disconnected correlation would not be a stable state.
\begin{table*}[htbp]
  \begin{center}
    \begin{small}
    \caption{The fitting results of $A_{\text{eff.}}$ and $\chi^{2}/\text{dof}$ of the effective mass and the estimated masses $m$}\label{tb:3}
    \begin{tabular}{|c|c|c|c|c|c|}\hline
      Eff. mass & $m_{c}$ & $A_{\text{eff.}}$  & $m$, MeV  & $FR$, $t/a$ & $\chi^{2}/\text{dof}$ \\ \hline
      $am_{s}$  & Normal conf  & 0.267(3) &  6.17(6)$\times10^{2}$ &  5--8 & 6/3 \\ \cline{2-6}
                          & 2 & 0.281(3) &  6.49(6)$\times10^{2}$  &  5--8 & 6/3 \\ \cline{2-6}
                          & 4 & 0.292(2) &  6.75(6)$\times10^{2}$  & 5--8 & 6/3 \\ \cline{2-6}
                          & 6 & 0.292(3) &  6.75(6)$\times10^{2}$  & 5--8 & 3/3 \\ \hline   
   $am_{\text{dis-}s}$  & Normal conf & 1.55(2)$\times10^{-2}$ & 35.8(0.5) & 2--4 & 22/2 \\ \cline{2-6}
                               & 2 & 2.10(4)$\times10^{-2}$ & 48.6(0.9) & 2--4 & 6/2 \\ \cline{2-6}
                               & 4 & 3.05(5)$\times10^{-2}$ & 70.6(1.1) & 2--4 & 7/2 \\ \cline{2-6}
                               & 6 & 3.16(6)$\times10^{-2}$ & 73.1(1.3) & 3--5 & 2/2 \\ \hline
    \end{tabular}
    \end{small}
  \end{center}
\end{table*}
\begin{figure*}[htbp]
  \begin{center}
    \includegraphics[width=163mm]{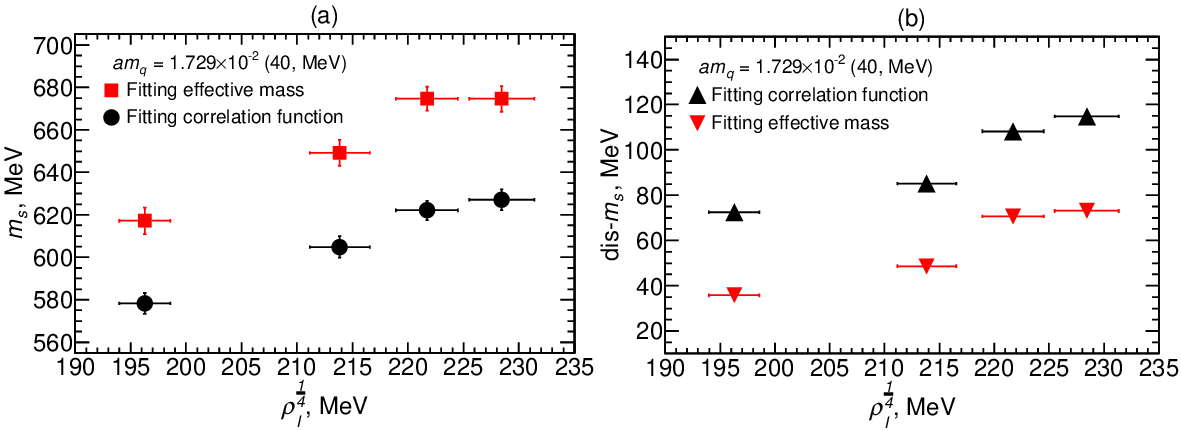}
  \end{center}
  \setlength\abovecaptionskip{-1pt}
  \caption{Comparisons of the masses obtained by fitting the curves to the numerical results of the correlations with the effective mass. (a) the masses $m_{s}$ of the connected contributions. (b) the masses $m_{\text{dis-}s}$ of the disconnected contributions.}\label{fig:3}
\end{figure*}

Figure~\ref{fig:3} shows the estimated masses $m_{s}$ and $m_{\text{dis-}s}$ from connected contributions and disconnected contributions. We compare the masses obtained by fitting curve~(\ref{eq:fitc}) to the correlation functions with the masses obtained from the effective mass~(\ref{eq:effemass}). The figure demonstrates that the masses $m_{s}$ and $m_{\text{dis-}s}$ increase with increasing the number density of the instantons and anti-instantons $\rho_{I}^{\frac{1}{4}}$.
\begin{table*}[htbp]
  \begin{center}
    \begin{small}
      \caption{The mass ratios. The denominator $m_{s}$ of $\frac{m_{s\text{-prime}}}{m_{s}}$ is the mass obtained by fitting the curve to the correlation function}\label{tb:4}
      \begin{tabular}{|c|c|c|c|c|}\hline
        $m_{c}$ & Corr. $\frac{m_{\text{dis-}s}}{m_{s}}$ & Eff. $\frac{m_{\text{dis-}s}}{m_{s}}$ & $\frac{m_{s\text{-prime}}}{m_{s}}$ & $\frac{m_{s\text{-prime}}}{m_{ss}}$ \\ \hline
        Normal conf &  0.125(5) &  5.80(10)$\times10^{-2}$ &  1.13(4) & 3.05(12) \\\hline 
        2 & 0.141(5) &  7.49(15)$\times10^{-2}$ &  1.18(4) & 3.16(12) \\\hline 
        4 & 0.174(4) &  0.1046(19)             &  1.36(4) & 3.35(10) \\\hline 
        6 & 0.183(4) &  0.108(2)               &  1.49(4) & 3.58(11) \\\hline 
      \end{tabular}
    \end{small}
  \end{center}
\end{table*}

To show the influences of the created instantons and anti-instantons on the masses estimated from the connected and disconnected correlations, we calculate the ratios $\frac{m_{\text{dis-}s}}{m_{s}}$ of the mass $m_{\text{dis-}s}$ of the disconnected contributions to the mass $m_{s}$ of the connected contributions. The calculated results are provided in Table~\ref{tb:4}.

\begin{figure*}[htbp]
  \begin{center}
    \includegraphics[width=165mm]{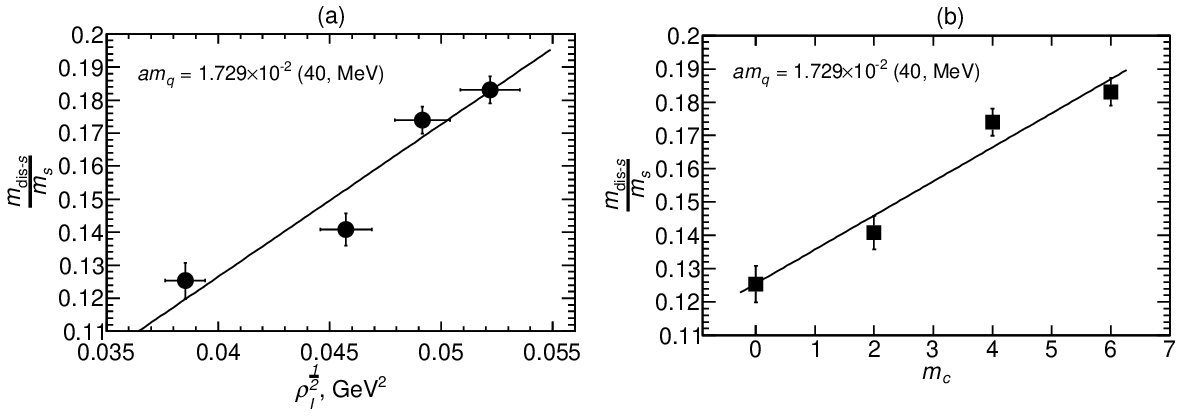}
  \end{center}
  \setlength\abovecaptionskip{-1pt}
  \caption{The mass ratios Corr. $\frac{m_{\text{dis-}s}}{m_{s}}$ and the fitting results. (b) the magnetic charge $m_{c}$ = 0 indicates the numerical results of the normal configurations.}\label{fig:4}
\end{figure*}

We then fit the following linear function to the numerical results of Corr. $\frac{m_{\text{dis-}s}}{m_{s}}$ in Table~\ref{tb:4} as shown in Fig.~\ref{fig:4}: $y = Ax + B$, $x = \rho_{I}^{\frac{1}{2}}$ or $x = m_{c}$. The fitting results are as follows: (i) $x = \rho_{I}^{\frac{1}{2}}$, $A$ = 4.6(7), $B$ = -6(3)$\times10^{-2}$, and $\chi^{2}/\text{dof}$ = 4/2. (ii) $x = m_{c}$, $A$ = 1.02(11)$\times10^{-2}$, $B$ = 0.126(4), and $\chi^{2}/\text{dof}$ = 5/2.

Figure~\ref{fig:4} and these fitting results indicate that the mass ratios $\frac{m_{\text{dis-}s}}{m_{s}}$ linearly increase with increasing the square root of the number density of the instantons and anti-instantons and the number of magnetic charges of the additional monopoles and anti-monopoles. These results indicate that the influences of the instantons and anti-instantons and the monopoles and anti-monopoles on the disconnected contributions are larger than the influences on the connected contributions.

Incidentally, the connected and disconnected correlations for the scalar density have already been estimated in~\cite{Isgur1}. Moreover, the instanton model predicts the contributions of the instantons to the connected and disconnected correlations for the scalar density in~\cite{Schafer1,Schafer2}.

\subsection{\textnormal{\textit{Scalar-Prime Meson Mass}}}

\renewcommand{\thefootnote}{\arabic{footnote}}

Finally, we use calculation methods regarding the eta-prime meson mass in lattice gauge theory~\cite{DeGrand3} as a reference and estimate the scalar-prime meson\footnote[1]{This is named without a theoretical background in this article for convenience.} mass from the disconnected correlation. We fit the following curve with two parameters $a^{6}Z_{\text{dis-}SS}$ and $am_{ss}$ supposing double poles to the numerical results of the disconnected correlation:
\begin{equation}
  C_{\text{dis-}SS}(t) = \frac{a^{6}Z_{\text{dis-}SS}}{(am_{ss})^{3}}\left\{(1 + m_{ss}t)\exp(-m_{ss}t) + \left[1 + m_{ss}(T - t)\right]\exp\left[-m_{ss}(T - t)\right]\right\}\label{eq:fit_func_discone}.
\end{equation}
The fitting results of $a^{6}Z_{\text{dis-}SS}$ and $am_{ss}$ are listed in Table~\ref{tb:5}. We then derive the square mass $m_{s\text{-prime}}^{2}$ of the scalar-prime meson from the following formula substituting the fitting results of $a^{4}Z$ in Table~\ref{tb:2}:
\begin{equation}
  a^{6}Z_{\text{dis-}SS} = \frac{a^{4}Z(am_{s\text{-prime}})^{2}}{4}\label{eq:fit_func_z2}.
\end{equation}
The numerical results of the masses of $m_{ss}$ and $m_{s\text{-prime}}$ are provided in Table~\ref{tb:5}. The calculated results of the mass ratios $\frac{m_{s\text{-prime}}}{m_{s}}$ and $\frac{m_{s\text{-prime}}}{m_{ss}}$ are listed in Table~\ref{tb:4}. 
\begin{table*}[htbp]
  \begin{center}
    \begin{small}
      \caption{The fitting results of $a^{6}Z_{\text{dis-}SS}$, $am_{ss}$, and $\chi^{2}/\text{dof}$ and the masses of $m_{ss}$ and $m_{s\text{-prime}}$}\label{tb:5}
      \begin{tabular}{|c|c|c|c|c|c|c|}\hline
        $m_{c}$ & $a^{6}Z_{\text{dis-}SS}$ & $am_{ss}$ & $m_{ss}$, MeV & $m_{s\text{-prime}}$, MeV & $FR$, $\Delta t/a$ & $\chi^{2}/\text{dof}$ \\ \hline
        Normal conf & 5.1(4)$\times10^{-5}$ & 9.24(17)$\times10^{-2}$ & 2.14(4)$\times10^{2}$ & 6.5(2)$\times10^{2}$ &  2--30 & 17/27 \\\hline   
                  2 & 8.3(5)$\times10^{-5}$ & 9.81(17)$\times10^{-2}$ & 2.27(4)$\times10^{2}$ & 7.2(2)$\times10^{2}$ &  2--30 & 25/27\\\hline 
                  4 & 1.45(7)$\times10^{-4}$ & 0.1091(14) & 2.52(3)$\times10^{2}$ & 8.5(2)$\times10^{2}$ &  2--30 & 47/27\\\hline              
                  6 & 1.79(9)$\times10^{-4}$ & 0.1129(14) & 2.61(3)$\times10^{2}$ & 9.3(2)$\times10^{2}$ &  2--30 & 41/27\\\hline              
      \end{tabular}
    \end{small}
  \end{center}
\end{table*}
\begin{figure*}[htbp]
  \begin{center}
    \includegraphics[width=165mm]{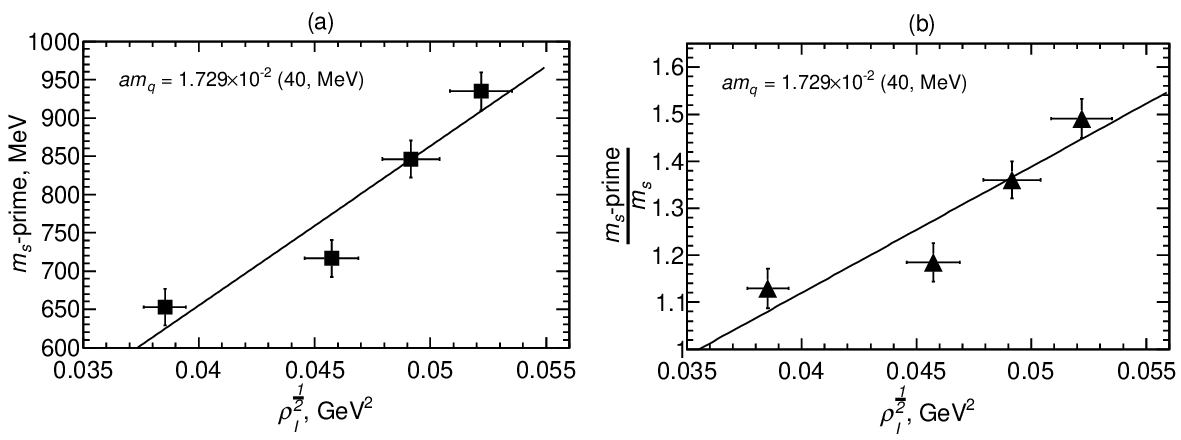}
  \end{center}
  \setlength\abovecaptionskip{-1pt}
  \caption{(a) the linear relationship between the scalar-prime mass $m_{s\text{-prime}}$ and the square root of the number density of the instantons and anti-instantons $\rho_{I}^{\frac{1}{2}}$. (b) the linear relationship between the mass ratio $\frac{m_{s\text{-prime}}}{m_{s}}$ and the square root of the number density of the instantons and anti-instantons $\rho_{I}^{\frac{1}{2}}$.}\label{fig:5}
\end{figure*}

To evaluate the increases in the mass $m_{s\text{-prime}}$, we fit the following linear function: $y = A_{s'}\rho_{I}^{\frac{1}{2}} + B_{s'}$ as shown in (a) of Fig.~\ref{fig:5}. The fitting results are as follows: $A_{s'}$ = 2.1(3)$\times10^{-2}$ Mev$^{-1}$, $B_{s'}$ = -1.8(1.5)$\times10^{2}$ MeV, and $\chi^{2}/\text{dof}$ = 4/2. These results demonstrate that the mass $m_{s\text{-prime}}$ becomes heavy with increasing the square root of the number density of the instantons and anti-instantons.

Moreover, we fit the following function to the numerical results of the mass ratio $\frac{m_{s\text{-prime}}}{m_{s}}$ as shown in (b) of Fig.~\ref{fig:5}: $y = A_{r}\rho_{I}^{\frac{1}{2}} + B_{r}$. The fitting results are $A_{r}$ = 27(5), $B_{r}$ = 0.05(0.2), and $\chi^{2}/\text{dof}$ = 5/2. These results demonstrate that the mass ratio $\frac{m_{s\text{-prime}}}{m_{s}}$ increases in direct proportion to the square root of the number density of the instantons and anti-instantons.

\section{\textnormal{Summary and conclusions}}\label{sec:3}

We have calculated the connected and disconnected correlation functions for the scalar density using the low-lying eigenvalues and eigenvectors of the overlap Dirac operator. We have investigated the effects of the monopoles, anti-monopoles, instantons, and anti-instantons that are created in the vacuum on the connected and disconnected correlations, and their effects on the masses that are estimated from these correlations.

The preliminary result shows that the masses that are estimated from the connected and disconnected correlations for the scalar density increase with increasing the number density of the instantons and anti-instantons. We have demonstrated that the mass ratio $\frac{m_{\text{dis-}s}}{m_{s}}$ of the disconnected contribution to the connected contribution linearly increases with increasing the square root of the number density of the instantons and anti-instantons and the number of magnetic charges of the additional monopoles and anti-monopoles.

Finally, we have estimated the scalar-prime meson mass from the disconnected contributions by fitting the same curve as defined in the estimation of the eta-prime meson mass in the quenched approximation. The preliminary results demonstrate that the scalar-prime meson mass and the mass ratio of the scalar-prime meson mass to the scalar meson increase by increasing the number density of instantons and anti-instantons. These increases are in direct proportion to the square root of the number density of instantons and anti-instantons.

These results indicate that the impacts of the instantons and anti-instantons and the monopoles and anti-monopoles on the disconnected contributions are larger than the influences on the connected contributions. We suppose that the additional monopoles and anti-monopoles would strongly affect the disconnected correlations from comparisons of our previous research. We are studying to reveal the reasons.

\section*{\textnormal{ACKNOWLEDGMENTS}}
The author uses the supercomputer SX-series and PC-clusters at the Research Center for Nuclear Physics (RCNP) and Cybermedia Center at Osaka University and the storage element of the Japan Lattice Data Grid at the RCNP. The author performed numerical calculations using XC40 at the Yukawa Institute for Theoretical Physics at Kyoto University. The author appreciates that computer resources are kindly provided for this research.

\end{document}